\documentclass[12pt, preprint]{aastex}

\def\lesssim{\mathrel{\hbox{\rlap{\hbox{\lower4pt\hbox{$\sim$}}}\hbox{$<$}}}}
\def\gtrsim{\mathrel{\hbox{\rlap{\hbox{\lower4pt\hbox{$\sim$}}}\hbox{$>$}}}}

\usepackage{graphicx}

\def\so{\Sigma_{\rm 0}}
\def\B{\begin{equation}}
\def\E{\end{equation}}
\def\Fo{F_{\rm 0}}
\def\B{\begin{equation}}
\def\E{\end{equation}}
\def\O{\Omega}

\def\ep{\epsilon}

\slugcomment{}
\shorttitle{Disks with Non-zero Inner Torque}
\shortauthors{Dezen \& Flores}

\begin{document}

\title{Spectra and Structure of Accretion Disks with Non-zero Inner Torque}

\shortauthors{Dezen \& Flores}

\author{Theodore Dezen and Bryan Flores}
\email{tdezen@sandiego.edu}
\affil{Department of Physics and Biophysics, University of San Diego, San Diego, CA 92110}

\begin{abstract}

We present numerical spectral and vertical structure calculations appropriate for near-Eddington luminosity, radiation pressure dominated accretion disks around stellar mass black holes. We cover a wide range of black hole spins, and incorporate dissipation profiles based on first-principles three-dimensional MHD disk interior simulations. We also include non-zero stresses at the ISCO, which results in the disk effective temperature to increase rapidly towards the black hole, and give rise to rather extreme conditions with high temperatures and low surface densities. We found that local annuli spectra become increasingly characteristic of saturated Comptonisation with decreasing distance to the black hole. While the spectra becomes harder with increasing black hole spin, they do not give rise to a broad power law tail even at maximum spin. We discuss the implications of our results in the context of the steep power law (SPL) state and the associated high-frequency quasi-periodic oscillations (HFQPO) observed in some X-ray binary systems.

\end{abstract}


\section{Introduction}

Galactic black hole X-ray binaries (BHB) show several states of outburst distinguished by luminosity, spectral shape and variability (see for example, \cite{mr06} and \cite{dgk07}). In particular, at their highest luminosities the spectra contains a steep power law component with photon index $\Gamma>2.4$ \citep{mr06}. These energetically significant power law tails begin at the spectral peak ($\approx10 \ \rm keV$) and could extend into the $MeV$ regime \citep{lw05, gr98}. Moreover, this steep power law (SPL) spectral state is accompanied by high-frequency ($\nu>50$ Hz) quasi-periodic oscillations (HFQPO) in the light curves when integrated over approximately $10$ to $30 \ \rm keV$ in photon energies.

Understanding the first-principles physics of radiating accretion flows that presumably underly these observational properties remain an important outstanding problem in astrophysics. The standard thin accretion disk model \citep{ss73, nt73, rh95} assumed that the stress and luminosity at the innermost stable circular orbit (ISCO) drops to zero, and that at this point the material essentially simply disappears into the black hole. This assumption received significant recent theoretical scrutiny upon the realization that magnetohydrodynamic turbulence \citep{bh91, bh98} is probably the source of stress that drives accretion. In particular, \cite{ak00} demonstrated that having non-zero magnetic stresses at the ISCO can cause the effective temperature to rise sharply towards the black hole instead of fall to zero as predicted by the standard model. Among the potentially observable consequences postulated by these authors, the inner disk consequently becomes effectively thin, and extends the spectrum to higher frequencies.

More recently, \cite{db14} (from here on referred to as DB14) proposed that the \cite{ak00} model provides feasible mechanism for explaining both the steep power law (SPL) state seen at near Eddington luminosities and the associated high frequency quasi-periodic oscillations (HFQPO). These authors argued that the rapidly rising effective temperature with decreasing distance to the black hole would give rise to the SPL spectra, while also providing a natural filter for the HFQPOs that do not require the entire disk to oscillate coherently. 

In this work, we undertake a detailed numerical study of the structure and spectra of near-Eddington accretion disks with non-zero magnetic stresses the ISCO, and particularly focus on the effects of black hole spin. Unlike previous efforts that relied on one-zone models, we self-consistently couple vertical structure to radiative transfer at each disk annuli, and generate spectra that fully incorporates effects Comptonisation and metal opacities. Our inputs are time and horizontally averaged vertical dissipation profiles from first-principles stratified shearing-box simulations of accretion flows \citep{hkb09}. These calculations evolve the time-dependent three-dimensional radiation magneto-hydrodynamic equations and accounts for the tidal vertical gravity from the black hole. In simulations over a wide range of box-integrated radiation to gas pressure ratios, the resulting vertical spatial dissipation profiles generally peak at around a pressure scale-height away from the disk mid-plane, and should capture the effects of MRI turbulence. Moreover, these simulations collectively indicate that the $\alpha$-prescription \citep{ss73} relationship between pressure and stress approximately hold \citep{hbk09}. This means we are justified, at least in light of recent simulations, to use the $\alpha$-model with modifications to account for non-zero inner torque to generate radial profiles of total surface density $\Sigma_0$ and effective temperature $T_{\rm eff}$ that are also necessary for our vertical structure and radiative transfer computations.

This paper is organized as follows. In section $2$ we outline our numerical methods, paying particular attention to how we incorporated non-zero inner torque. Section $3$ showcases our numerical results, including full-disk spectra for all black hole spin values we covered. We turn to the possibility of HFQPOs in section $4$, and conclude in section $5$ with a discussion of on-going and future work.

\section{Methods}

We compute one-dimensional, time-independent and non-LTE disk annuli vertical structure and spectral models using the stellar atmosphere code TLUSTY \citep{hl95} that has been adapted for accretion disk applications \citep{h98, h00}. The program self-consistently solves the disk vertical structure equations along with angle-dependent and multi-frequency radiative transfer, and does not assume that the spectrum is modified from a blackbody via spectral hardening factors.  We treat electron (both Thomson and Compton) scattering with an angle-averaged Kompaneets source function \citep{h01}, and allow for deviations from local thermodynamic equilibrium by explicitly computing ion populations for hydrogen, helium and metals. 

\begin{figure}[h]
\includegraphics[width=12cm]{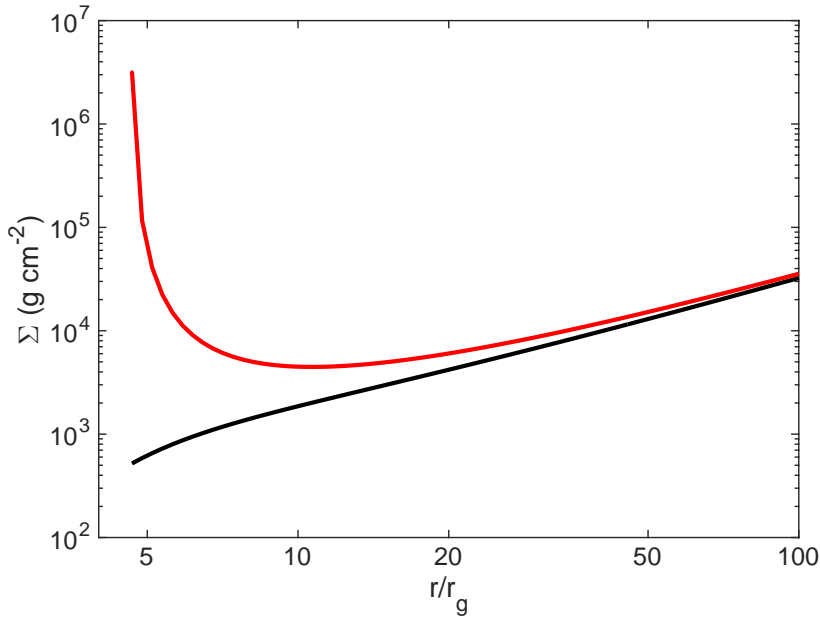}
\caption{Total disk surface density $\Sigma_0$ as a function of distance from black hole for $a/M=0.4$. Note that surface density continues to decrease towards the ISCO when $\Delta\ep=0.1$ (black) but rises sharply for the disk with $\Delta\ep=0$ (red).}
\label{fig:sigma}
\end{figure}

\begin{figure}[h]
\includegraphics[width=12cm]{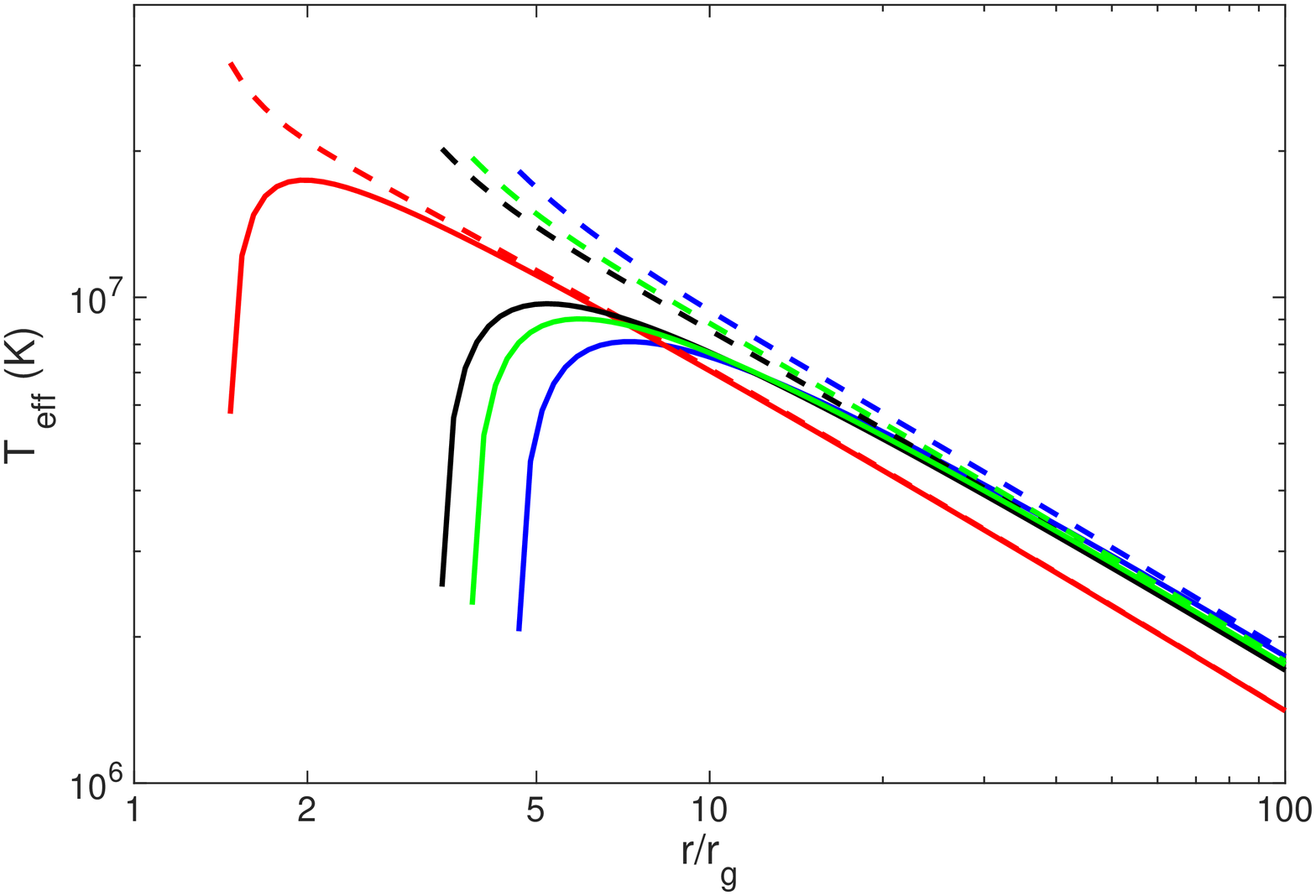}
\caption{Representative radial effective temperature profiles as a function of distance from the black hole normalized to the gravitational radii for accretion disks with zero (solid) and non-zero inner torque (dotted), respectively. The colors correspond to models with different black hole spins: blue, $a/M=0.4$; green, $a/M=0.6$; black, $a/M=0.7$; red, $a/M=0.99$. Note that the effective temperature continues to rise sharply towards the black hole all the way down to the inner edge for the black curve.}
\label{fig:profiles}
\end{figure}

TLUSTY self-consistently calculates the vertical structure of and local emergent spectrum from an accretion disk annuli. To specify an annuli, TLUSTY requires as input the mid-plane surface density $\Sigma_0$ (Figure \ref{fig:sigma}), effective temperature $T_{\rm eff}$ (Figure \ref{fig:profiles}), and orbital frequency squared $\Omega^2$, all of which are functions of distance from the black hole. Interested readers can consult previous TLUSTY-based work such as \cite{dav05} and \cite{tb13} for more details on using and modifying the code, as well as further discussions of numerical considerations. Next, we focus on how we generated these input parameters for a given blackhole mass and spin while incorporating both non-zero torque at disk inner boundary and simulation based dissipation physics.

We made several assumptions and approximations may impact the disk spectra and vertical structure that we will compute. First, we characterized the ionized plasma with a single gas temperature $T_{\rm g}$ because the mean free path of electron-ion collisions is much less than disk scale height, and therefore the electrons and protons are thermally locked. Further, we only consider vertical radiative diffusion for energy transport, so that we can relate the dissipation rate $Q$ to the frequency-integrated flux $F$ via
\B
Q(z)=\frac{dF}{dz},
\label{qdfdz}
\E
where $z$ is the height measured vertically upward (with $z=0$ at mid-plane). By analyzing several shearing box simulations, \cite{bla11} found that radiative diffusion is probably not the only important vertical energy transport mechanism. Specifically, radiation pressure work and magnetic buoyancy driven radiation advection can be as important as diffusive flux near the mid-planes of radiation pressure dominated disks. However, these simulations also showed that radiative diffusion still dominates at higher altitudes, so that in the end it is still the primary means to carry dissipated energy into the photospheric region where the spectrum forms. We therefore suspect that the emergent spectra will not change drastically even if we included radiative advection. On the other hand, more recent global simulations found radiative advection to be important even in the outflow regions of super-Eddington accretion flows \citep{jia14}, but how these results would impact the spectra is not yet clear. Finally, we ignore potentially important the effects of irradiation and outflow.

We also do not include magnetic fields in this study. While simulations such as those described in \cite{hkb09} found that magnetic pressure can be just as important, and even dominate over, radiation pressure in disk outer layers. However, \cite{dav09} showed that the expected hardening of the spectra due to significant magnetic support against gravity is largely compensated by the softening effects of complex density inhomogeneities around the photospheric regions. Moreover, radiative transfer calculations incorporating simulation-based vertical magnetic acceleration \citep{tb13} showed that dissipation profiles is what primarily drove significant changes in spectral shapes.

To generate input for TLUSTY and study the effects of non-zero inner torque, we use the results of \cite{ak00} as implemented by DB14 to numerically compute the vertically integrated surface density $\Sigma_0$ (Figure \ref{fig:sigma}), effective temperature $T_{eff}$ (Figure \ref{fig:profiles}) and orbital frequency $\O$ as a function of distance from the black hole by solving one-zone relativistic radial structure equations. Compared to the standard thin disk models \citep{nt73}, \cite{ak00} adds a single parameter $\Delta\ep$ that characterizes the additional radiative efficiency (relative to one computed in terms of only the binding energy) due to including a non-zero torque at the disk's inner edge. Our modeled disks accrete onto a $7$ solar mass black hole with $\alpha=0.02$  \citep{ss73}. We chose the normalized accretion rates
\B
\dot{m}=\frac{\ep\dot{M}c^2}{L_{\rm Edd}}=\frac{(\ep_0+\Delta\ep)\dot{M}c^2}{L_{\rm Edd}}
\E 
such that the disk luminosity is approximately Eddington, where $\ep_0$ is the accretion efficiency without enhancement due to the inner torque. Our models cover several black hole spins ranging from $a/M=0.4$ to $a/M=0.99$. For all spin values, we set $\Delta\ep$ to $0.1$. We divide each disk into many (on the order of $50$, and vary slightly between models) annuli starting at the ISCO. Finally, we use a relativistic transfer code \citep{agol97} to obtain the full disk spectra as seen by distant observers.

\begin{figure}[h]
\includegraphics[width=12cm]{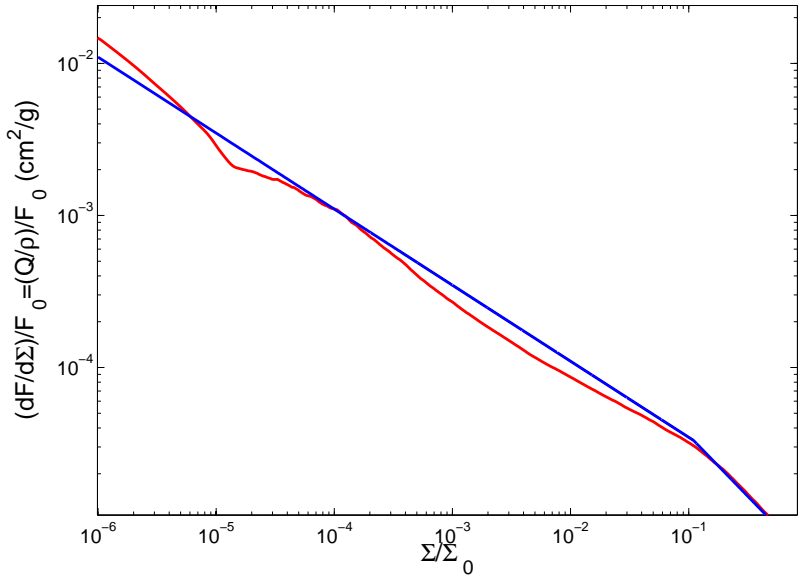}
\caption{Dissipation per unit mass normalized to total flux as a function of fractional surface density for a representative accretion disk annulus. The blue curve denotes the broken-power law expression of equation (\ref{dis1}), while the red curve is the horizontal and time averaged dissipation profile from simulation $1112a$ of \cite{hkb09}. While equation (\ref{dis1}) gives $Q\Sigma/\rho$, here we instead graphed $Q/\rho$ to emphasize the dissipation per unit mass in the upper layers.}
\label{fig:q1112a}
\end{figure}

We incorporate simulation-based dissipation physics by writing the vertical dissipation profile as a function of fractional surface density. Specifically, we fit the horizontal and time averaged dissipation profile from \cite{hkb09} using the broken power-law expression \citep{tb13}
\B
\frac{Q\Sigma}{\rho}=-\Sigma\frac{dF}{d\Sigma}=F_0\left\{\begin{array}{ccc}A\left(\frac{\Sigma}{\so}\right)^{0.5}&,& \Sigma/\so<0.11\\B\left(\frac{\Sigma}{\so}\right)^{0.2}&,& \Sigma/\so>0.11\end{array}\right..
\label{dis1}
\E
as shown in Figure \ref{fig:q1112a}, where the power-law indices and the $\Sigma/\Sigma_0=0.11$ division point are all required input for TLUSTY. Here $\Fo$ (obtained from $T_{\rm eff}$) and $\so$ are the total emergent flux and mid-plane surface density, respectively.  The code automatically computed the normalization constants $A (\approx 0.65)$ and $B (\approx 0.33)$ to ensure that the profile is continuous at the division point, and that the total flux is $F_0$ so that
\B
F_0=\int_0^{\Sigma_0}\frac{dF}{d\Sigma} \ d\Sigma.
\E
The surface density $\Sigma$ at height $z$ is given by
\B
\Sigma(z)=\int_{z}^{\infty}\rho(z')dz',  
\E 
where $\rho$ is the depth dependent density. We define $z=0$ to be the disk mid-plane, and hence $\rho$ and $\Sigma$ both go to zero at infinity. In a previous TLUSTY calculation, \cite{dav05} used a qualitatively similar prescription with a power law index of $0.5$ in the entire vertical disk domain. Here we fit the simulation in more detail, resulting in the break at $\Sigma/\Sigma_0=0.11$. This dissipation profile presumably captures the effects of turbulence due to the magnetorotational instability \citep{bh91}.

Before moving on, we note that a slim \citep{ab88}, rather than thin, disk model is likely more appropriate for high luminosity systems radiating at near or beyond the Eddington limit.  Moreover, we ignored the likely important effects of radial advection associated with such slim disks, and computed the radial profile essentially from height-averaged quantities. Properly incorporating these effects requires two-dimensional slim disk models that couple the radial and vertical structures, such as those proposed in \cite{sa11}. At present, these models do not yet include self-consistent multi-frequency radiative transfer. 

\section{Numerical Results}

\subsection{Annuli Spectra and Vertical Structure}

\begin{figure}
\includegraphics[width=9cm]{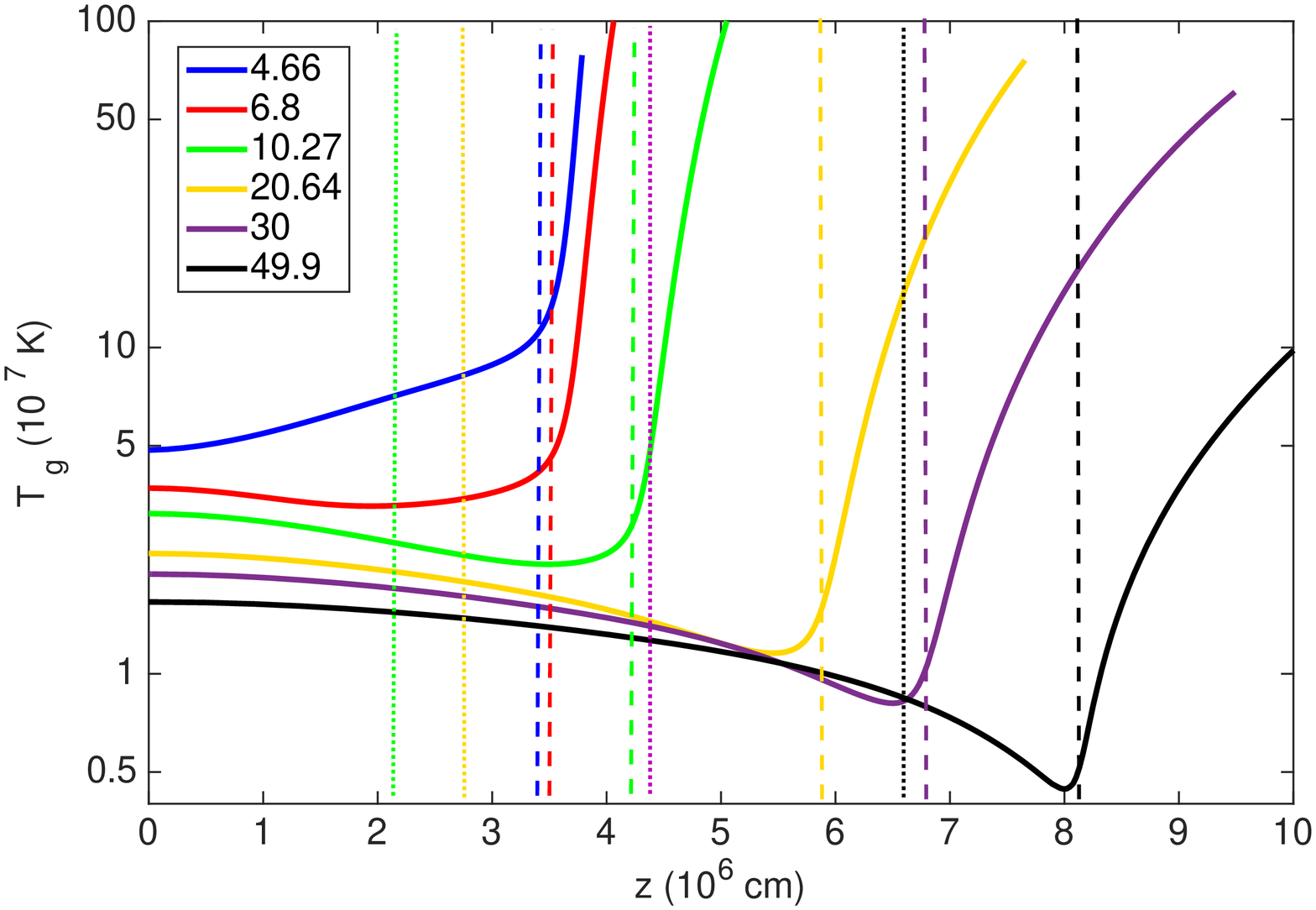}
\includegraphics[width=9cm]{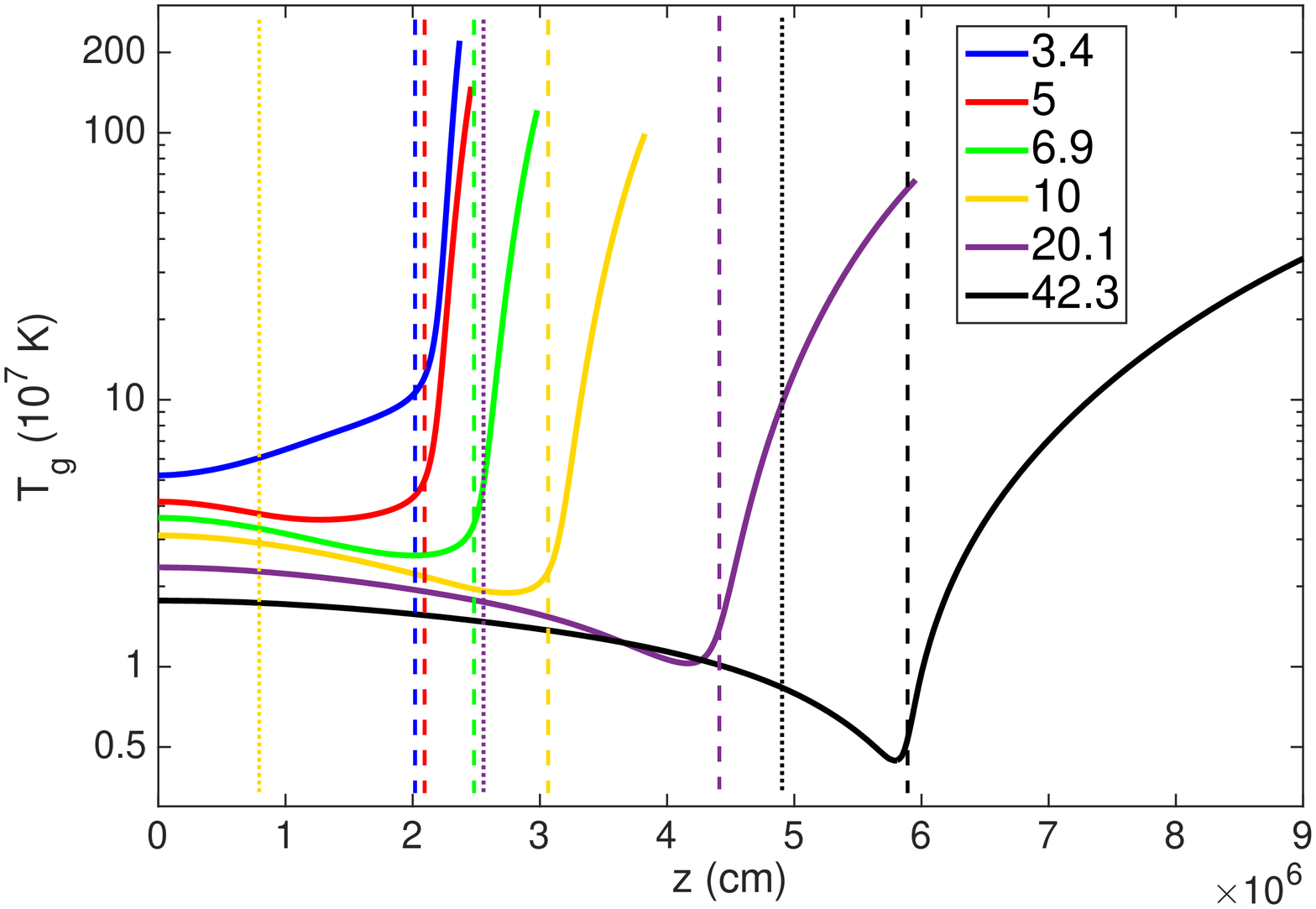}
\includegraphics[width=9cm]{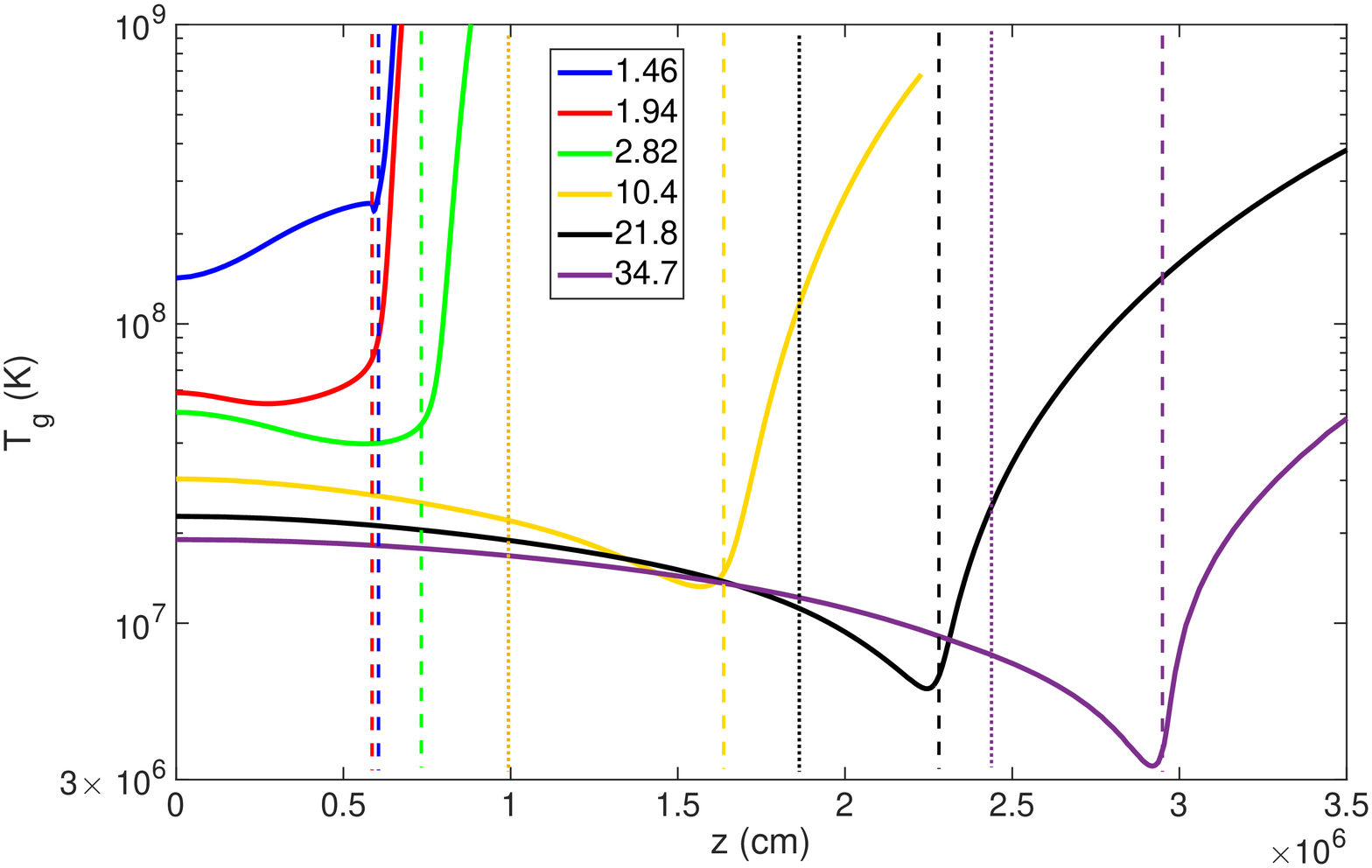}
\caption{Sample annuli gas temperature $T_{\rm g}$ versus height $z$ for the $a/M=0.4$ (top left panel), $0.7$ (top right panel) and $0.99$ (bottom panel) disks. The numbers in the legend are annuli radii in units of $r/r_{\rm g}$. The vertical dashed and dotted lines indicate the locations of scattering and effective photospheres, respectively. Note that the annuli become effectively thin but remain scattering thick near the black hole, as illustrated by the models with $r/r_{\rm g}<10$ on this figure.}
\label{fig:tg}
\end{figure}

\begin{figure}
\includegraphics[width=9cm]{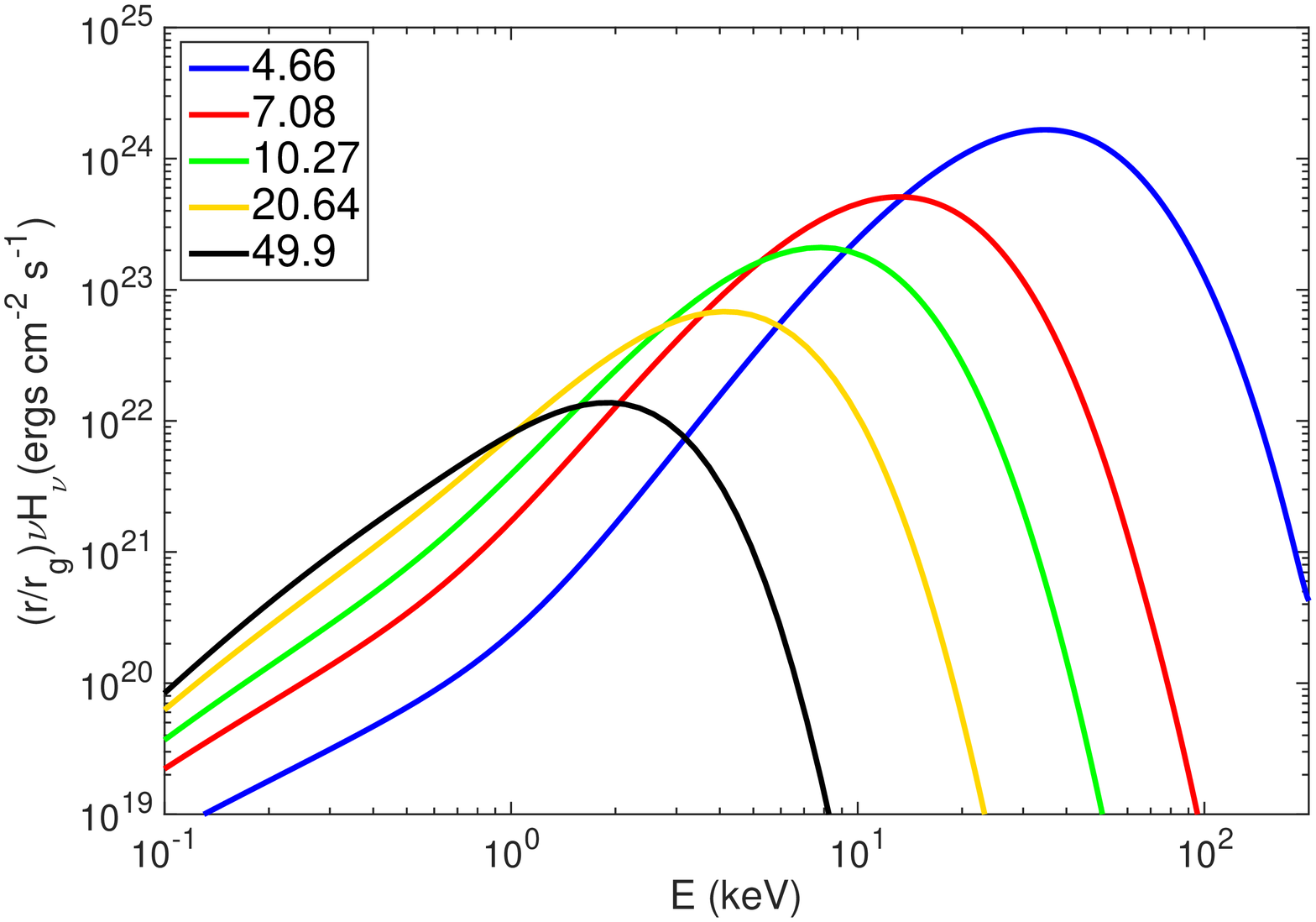}
\includegraphics[width=9cm]{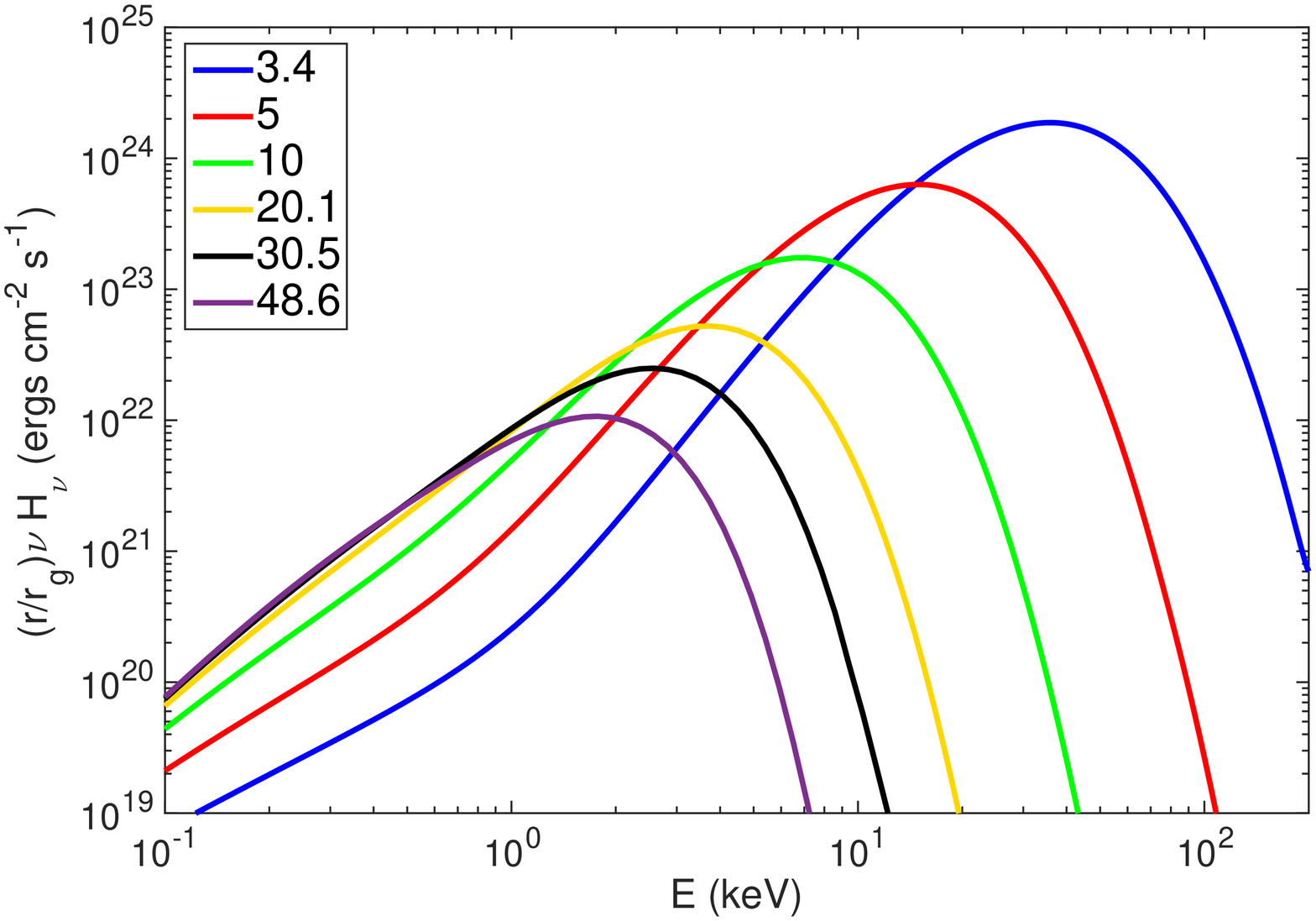}
\includegraphics[width=9cm]{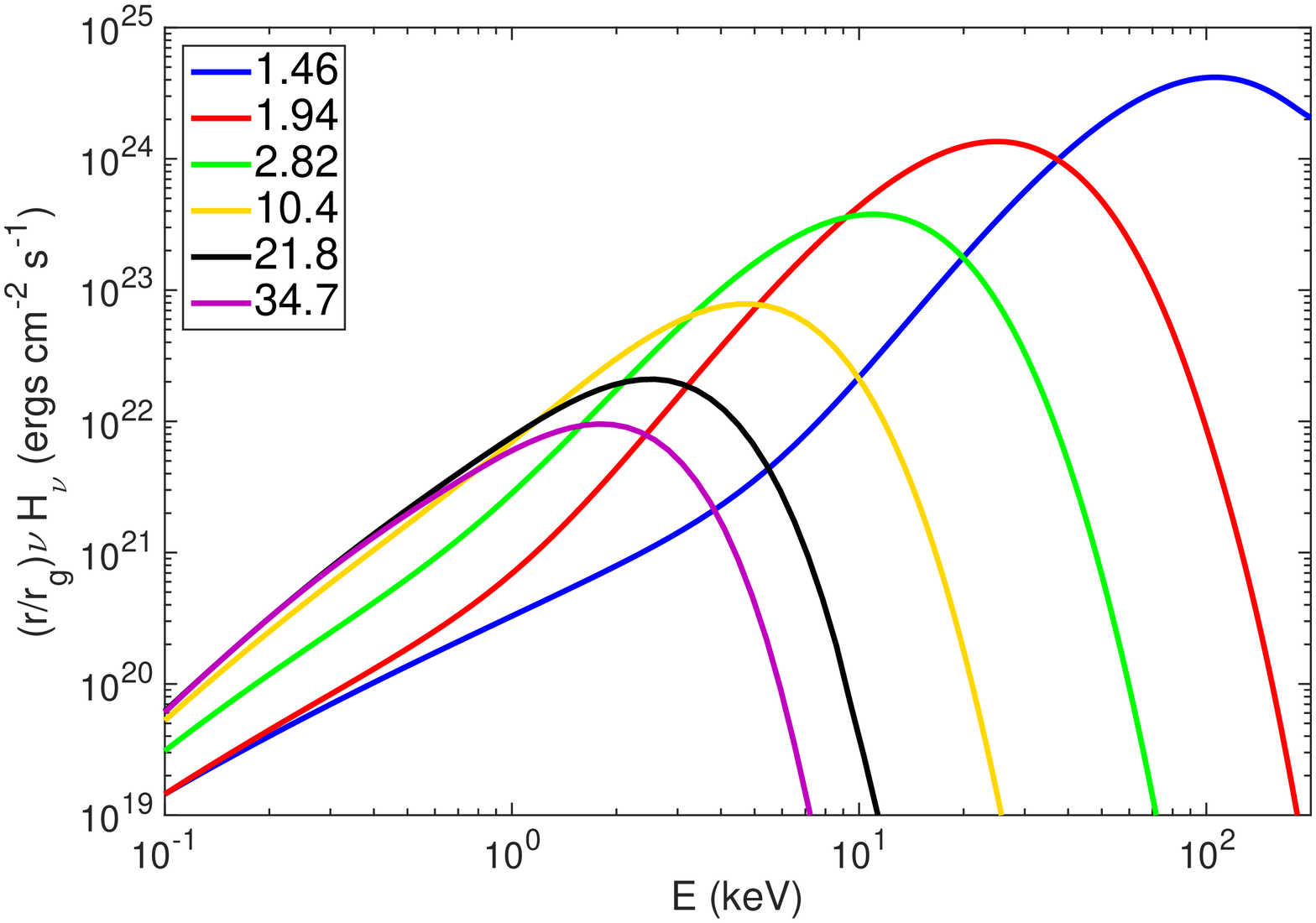}
\caption{Representative area weighted local annuli spectra for the disks with $a/M=0.4$ (upper left panel), $0.7$ (upper right) and $0.99$ (lower). The numbers in the legend denote annuli radii in units of $r/r_{\rm g}$. For these models, as $r/r_g$ decreases, the effective temperature increases and surface density decreases, resulting in spectral transitions from modified black body to saturated Compton scattering.}
\label{fig:annulispec}
\end{figure}

\begin{figure}[h]
\includegraphics[width=12cm]{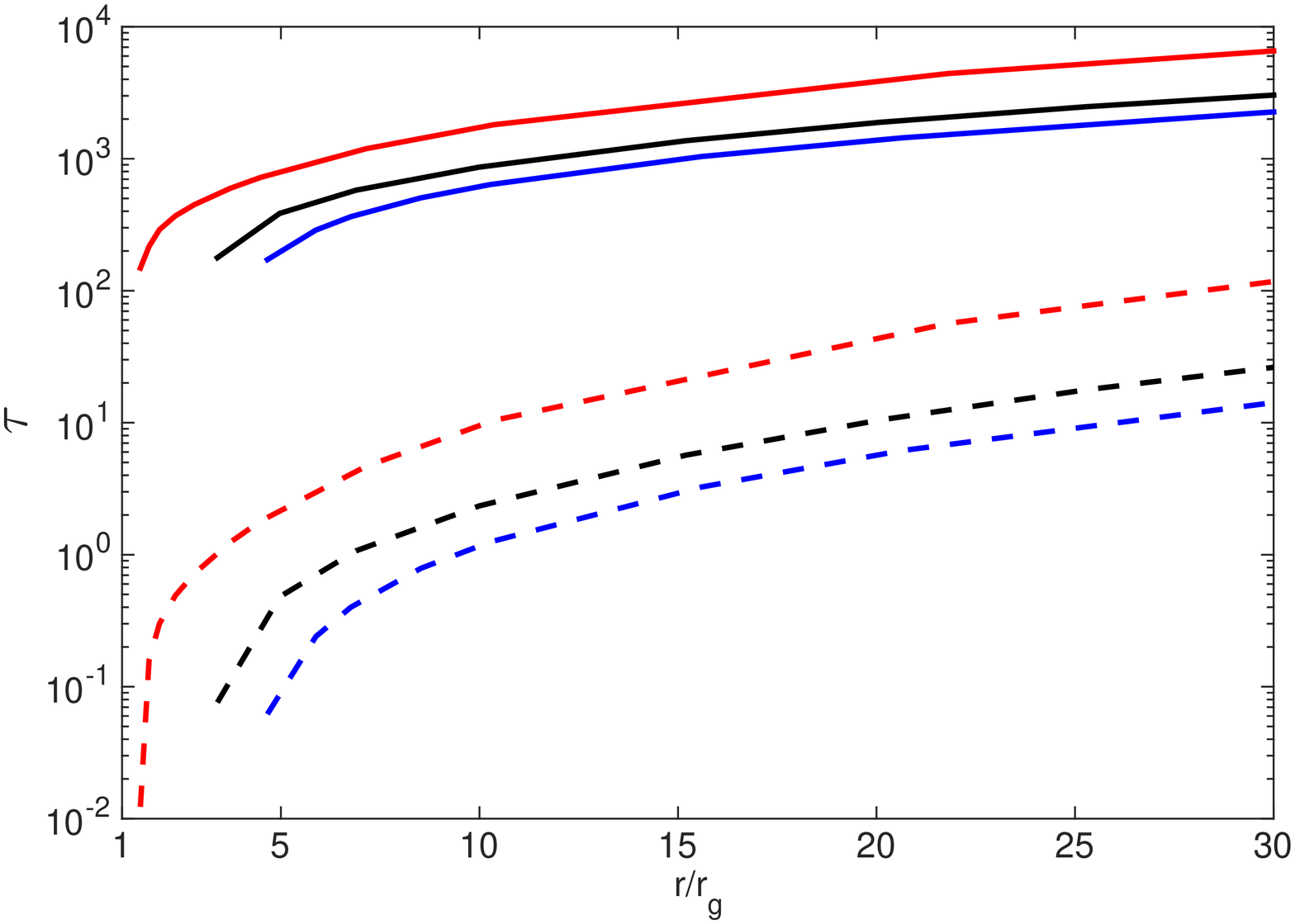}
\caption{Total vertical optical depth as a function of $r/r_{\rm g}$ for the $a/M=0.4$ (blue), $a/M=0.7$ (black) and $a/M=0.99$ (red) disks. The solid and dashed curves correspond to effective and scattering optical depths, respectively. Compared to standard \citep{nt73} thin disks, the optical depth plunges instead of increases as we approach the black hole.}
\label{fig:tau}
\end{figure}

\begin{figure}[h]
\includegraphics[width=12cm]{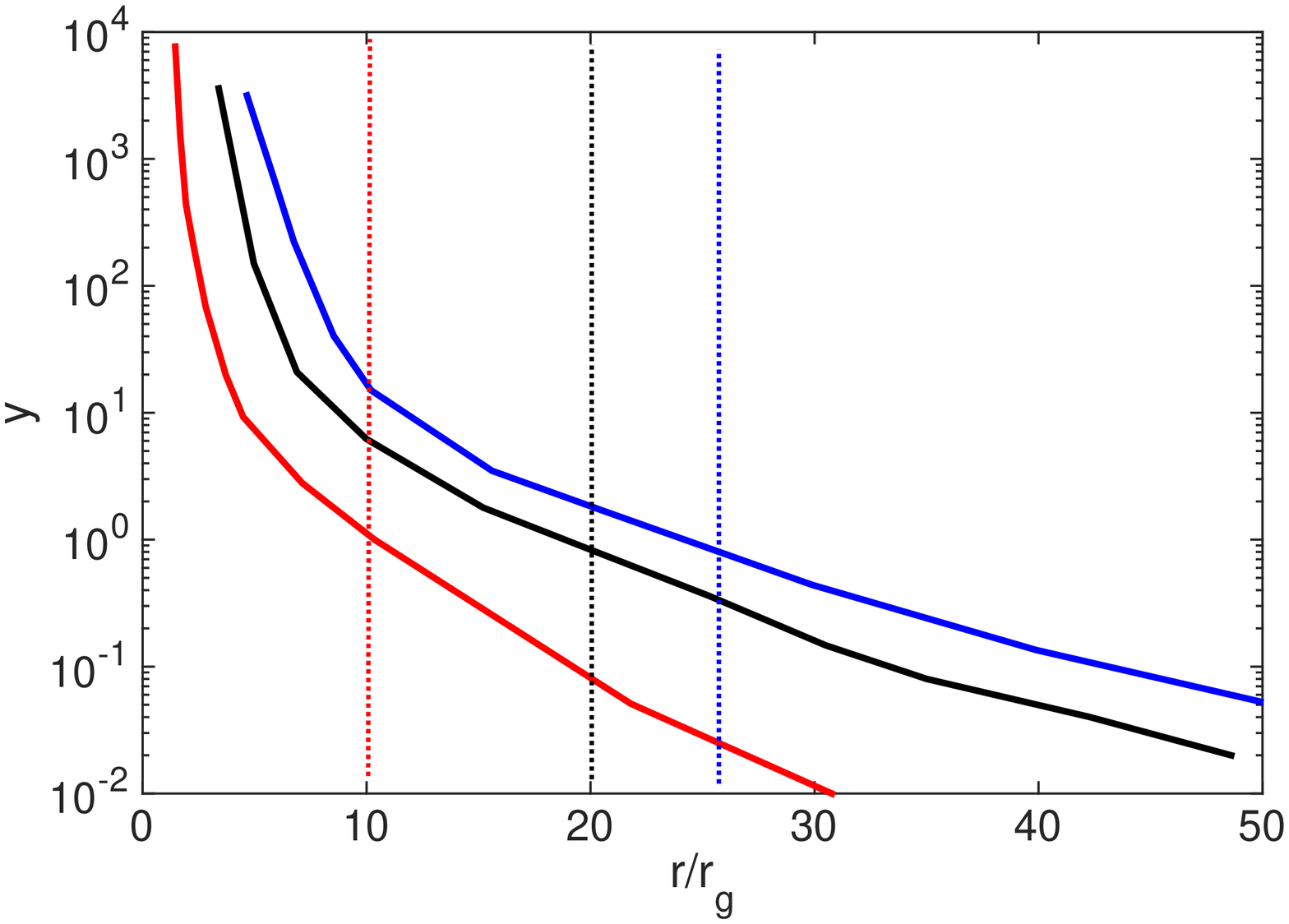}
\caption{Representative Compton $y$-parameters as functions of $r/r_g$ for disks with different spins. Here blue, black and red denote $a/M=0.4$, $0.7$ and $0.99$, respectively. The vertical dotted lines denote the annuli where the total effective optical depth is about 10, which correspond approximately to $y\approx1$.}
\label{fig:y}
\end{figure}

Figures \ref{fig:tg} and \ref{fig:annulispec} show representative vertical temperature profiles and emergent spectra, respectively, from several annuli models from annuli with $a/M=0.4$, $0.7$ and $0.99$. For all black hole spins, at large (tens of $r_{\rm g}$) radii the annuli are both effectively and scattering thick, resulting in modified blackbody spectra. As shown in Figure \ref{fig:tau}, closer to the black hole the disk becomes effectively thin but not to the extent predicted by one-zone calculations of DB14. Nonetheless, even within $10r_{\rm g}$, the disk remain very much optically thick to scattering, and the local annuli spectra transition to Wien shape, characteristic of saturated Comptonisation. To further illustrate this transition, we computed vertically integrated Compton $y$-parameters for each annuli starting from the gas temperature minima, where radiation and gas temperatures begin to decouple. From this point upward (towards the disk surface), photons may be up-scattered by higher energy electrons. This process continues until the scattering photosphere, where our integration ends. Figure \ref{fig:y} shows representative $y$-parameter results, all of which begin at much lower than $1$ at large radii but increase to much larger than $1$ closer to the black hole. 

Note that even the innermost annuli for disks around higher spin ($0.8$ and $0.99$) black holes do not show powerful high-energy power-law spectral tails characteristic of unsaturated Compton up-scattering. This is because while the gas temperature in the photospheric regions can reach beyond $10^8 \ \rm K$, none of our annuli models show a rapid and large increase between the temperature minima and the scattering photosphere, making it impossible to significantly up-scatter photons within a small number of collisions before they lose thermal coupling with the gas. These results qualitatively agree with previous disk atmosphere and radiative transfer calculations \citep{dav05, tb13} that found producing strong power-law annuli spectra requires additional dissipation in the spectral-forming regions than what we obtained by fitting vertical dissipation profiles from the \cite{hkb09} simulations. Incorporating torque at the inner disk edge, on the other hand, modifies the radial surface density and effective temperature profiles but does not change the vertical distribution of accretion power dissipated within each annuli model. 

\subsection{Full-Disk Spectra}

\begin{figure}
\includegraphics[width=12cm]{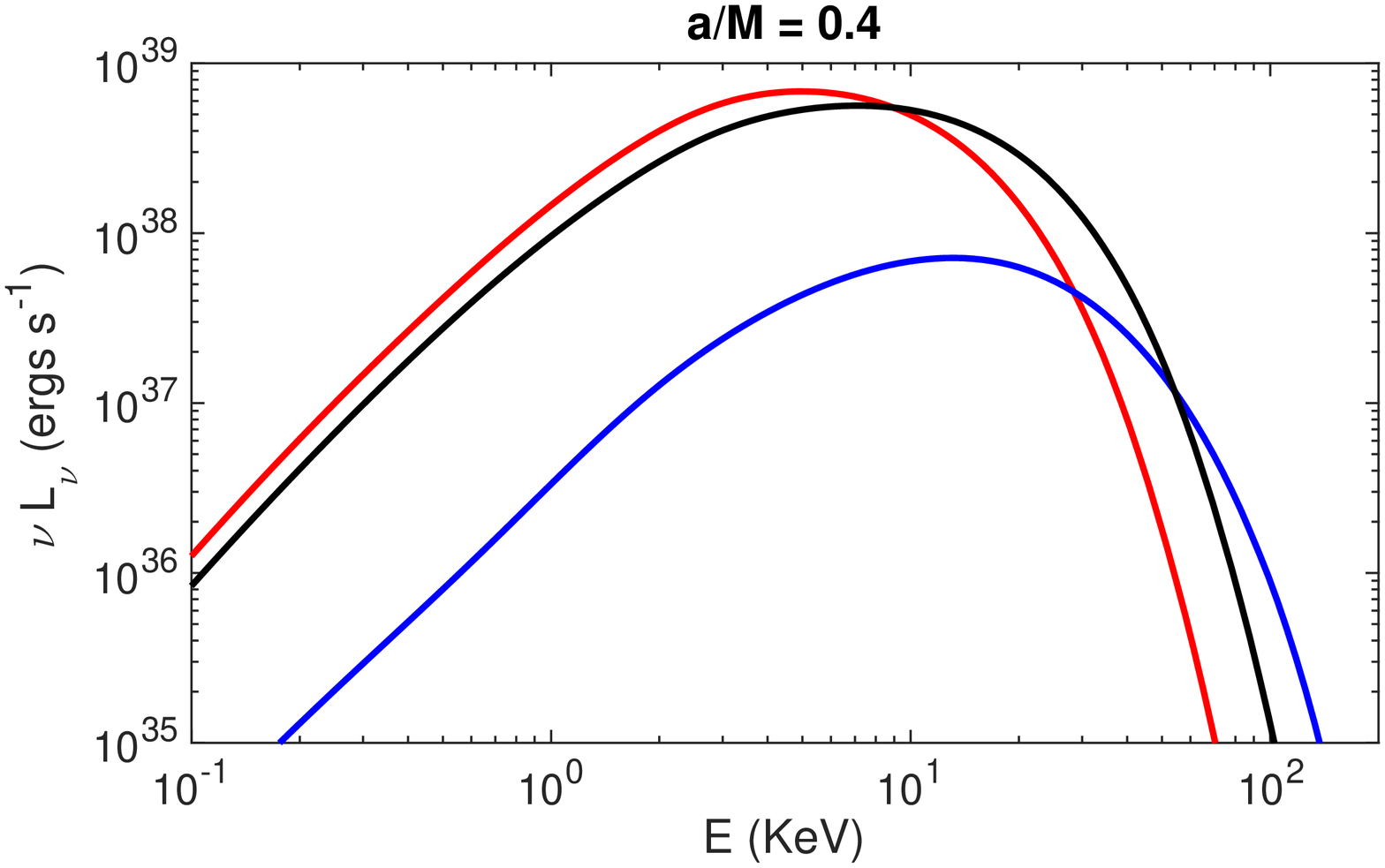}
\includegraphics[width=12cm]{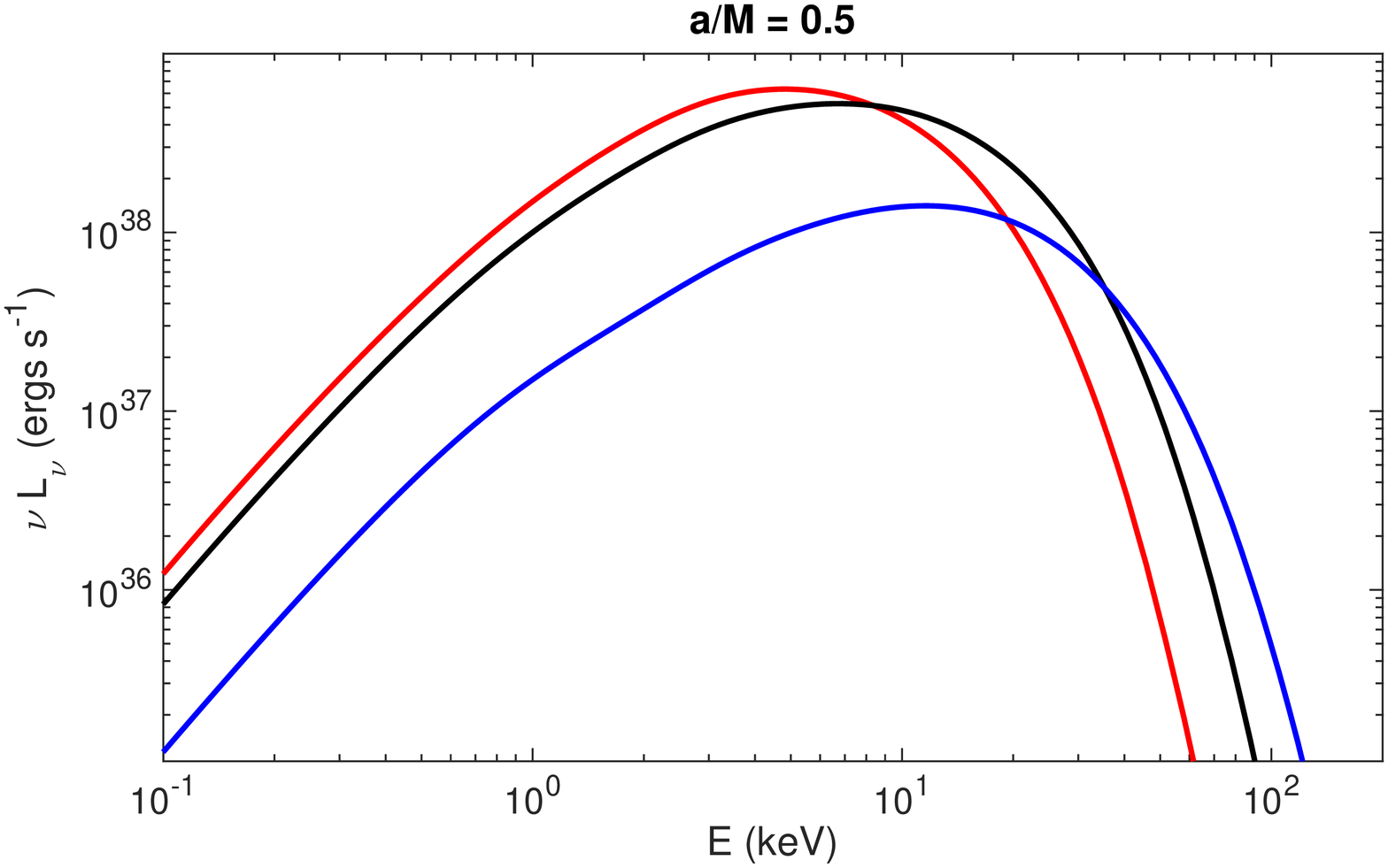}
\caption{Disk integrated spectra as seen by distant observers for disks with lower $a/M$ values. The red, black and blue curves correspond viewing the disk face-on, at $\theta=\pi/4$ (relative to the direction normal to disk surface) and nearly edge-on, respectively. While non-thermal, these spectra do not exhibit energetic high-energy tails characteristic of the SPL state.}
\label{fig:integrated}
\end{figure}

\begin{figure}
\includegraphics[width=12cm]{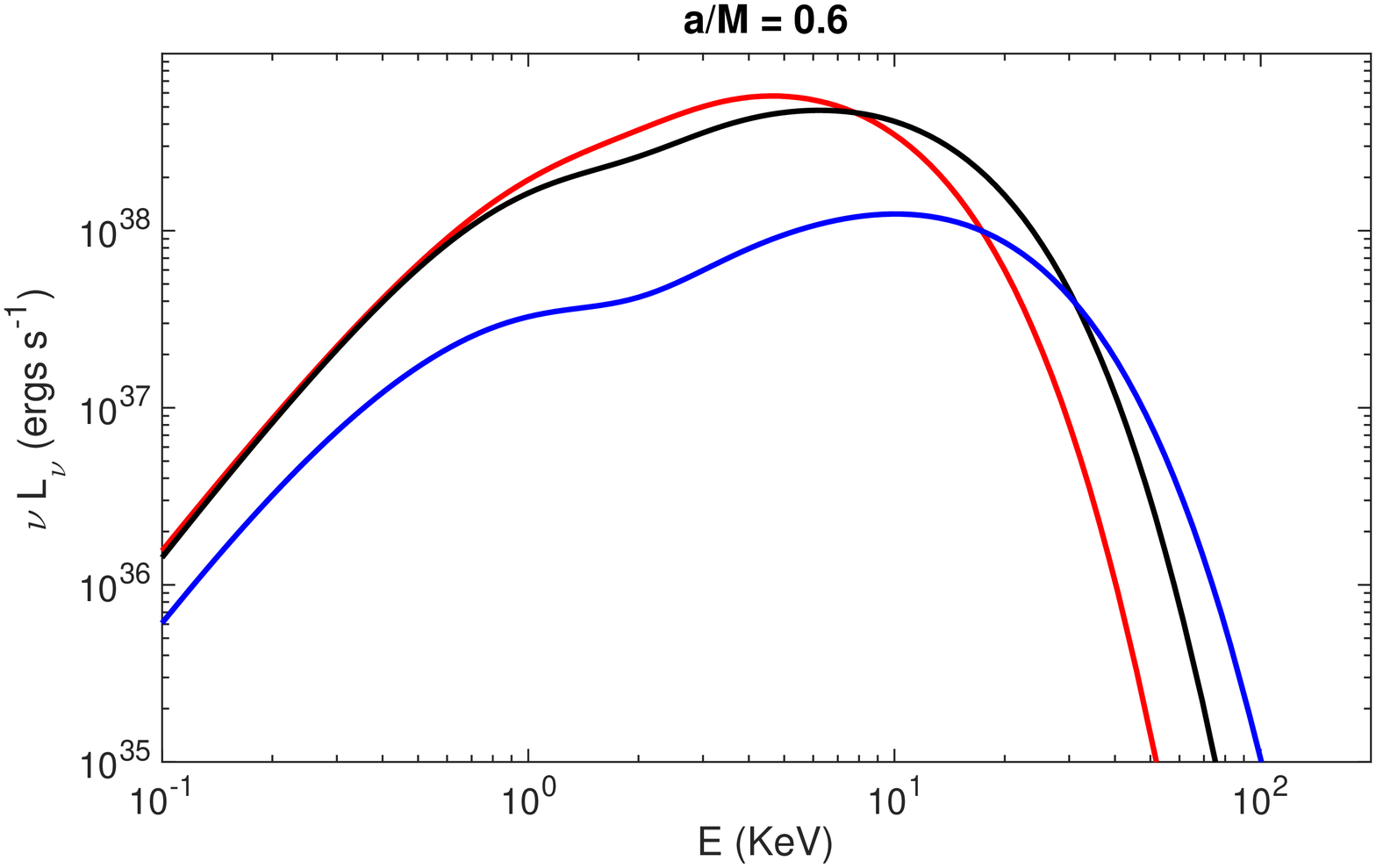}
\includegraphics[width=12cm]{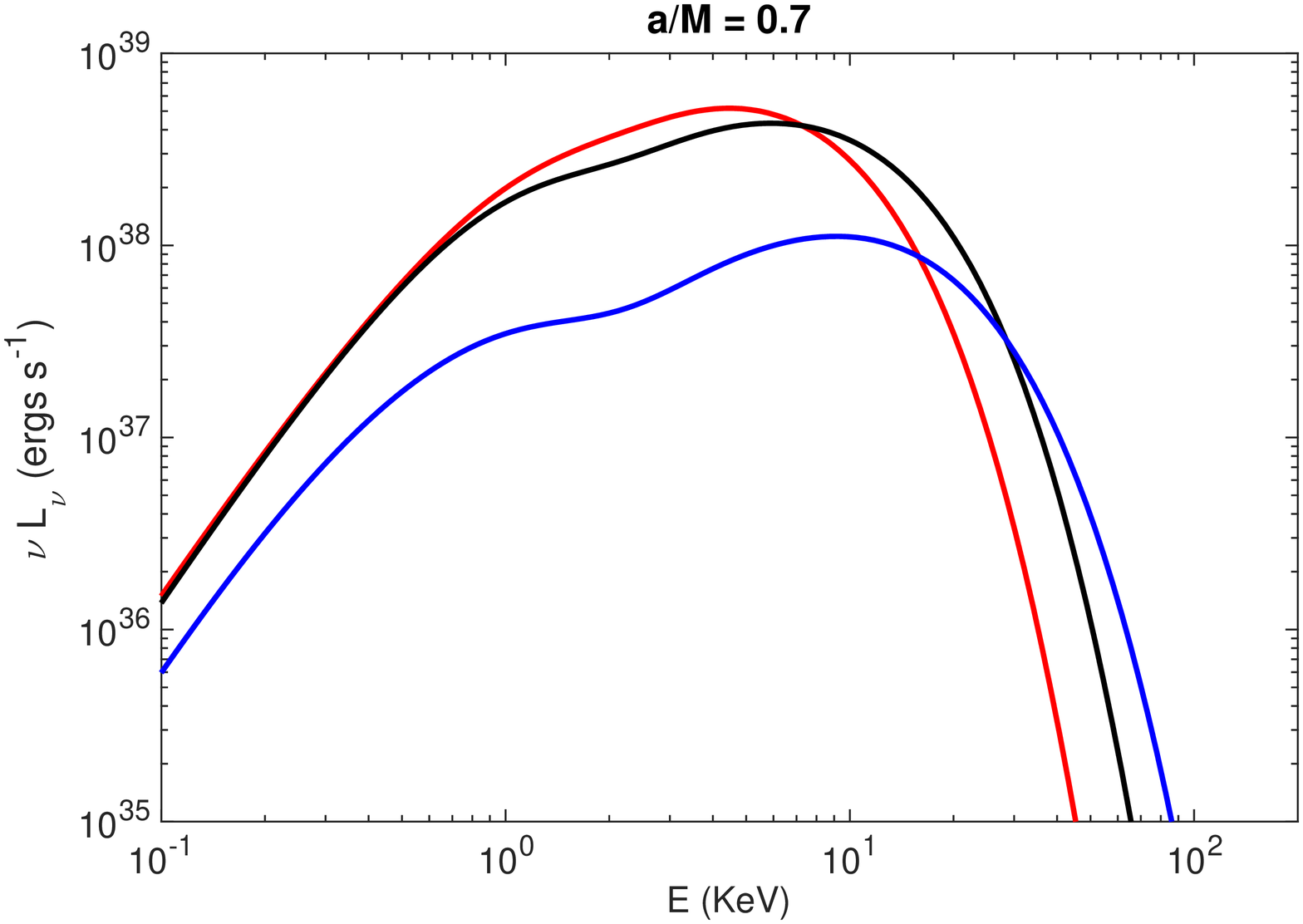}
\caption{Disk integrated spectra as seen by distant observers for disks with moderate $a/M$ values. The red, black and blue curves correspond viewing the disk face-on, at $\theta=\pi/4$ (relative to the direction normal to disk surface) and nearly edge-on, respectively. Compared to $a/M=0.4$ and $0.5$ models, these disk display a broad dip-like feature between $1$ and $10 \rm keV$), which becomes more prominent at higher viewing angles.}
\label{fig:intmid}
\end{figure}

\begin{figure}
\includegraphics[width=12cm]{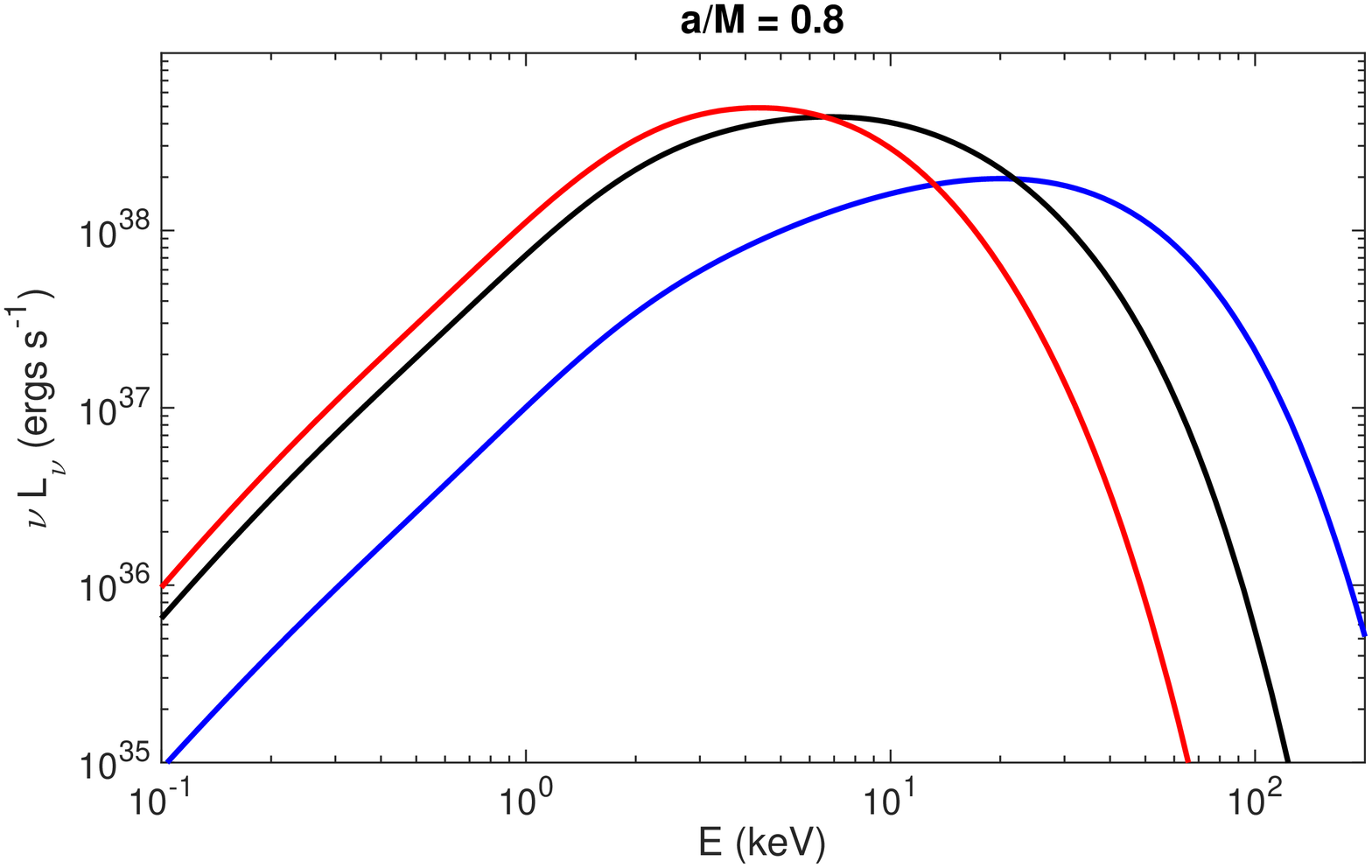}
\includegraphics[width=12cm]{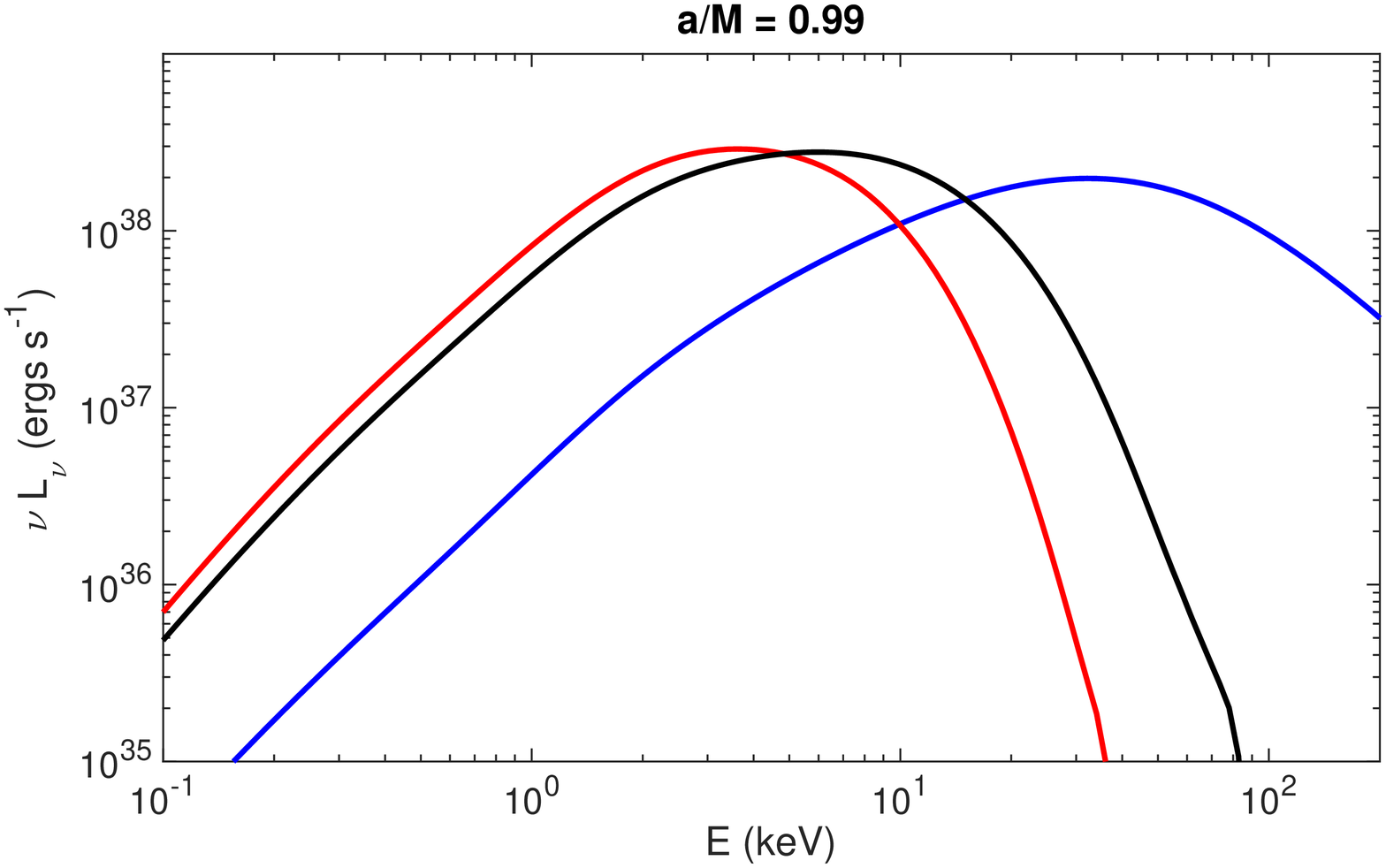}
\caption{Disk integrated spectra as seen by distant observers for disks with high $a/M$ values. Again, the red, black and blue curves correspond viewing the disk face-on, at $\theta=\pi/4$ and nearly edge-on, respectively. Note that relativistic effects are so strong in the $a/M=0.99$ case that the spectra does show a powerful high-energy tail when viewed nearly edge-on, although it still does not resemble SPL observations}
\label{fig:inthigh}
\end{figure}

Figure \ref{fig:integrated} shows the full-disk spectra from the $a/M=0.4$ and $a/M=0.5$ models. At these lower spins, the spectra is harder than one that would result from purely thermal local emission due to the modified black body contributions from the majority of annuli. As we move to higher spins of $a/M=0.6$ and $0.7$, the full-disk spectra (Figure \ref{fig:intmid}) broadens immediately red-ward of the peak as a result of more contributions from annuli spectra that thermalized due to strong Comptonisation.  For more extreme spins ($a/M=0.8$ and $0.99$, Figure \ref{fig:inthigh}), the spectra is more thermal as the fraction of total emitted power contributed by annuli exhibiting saturated Comptonisation further increases.  

\begin{figure}
\includegraphics[width=12cm]{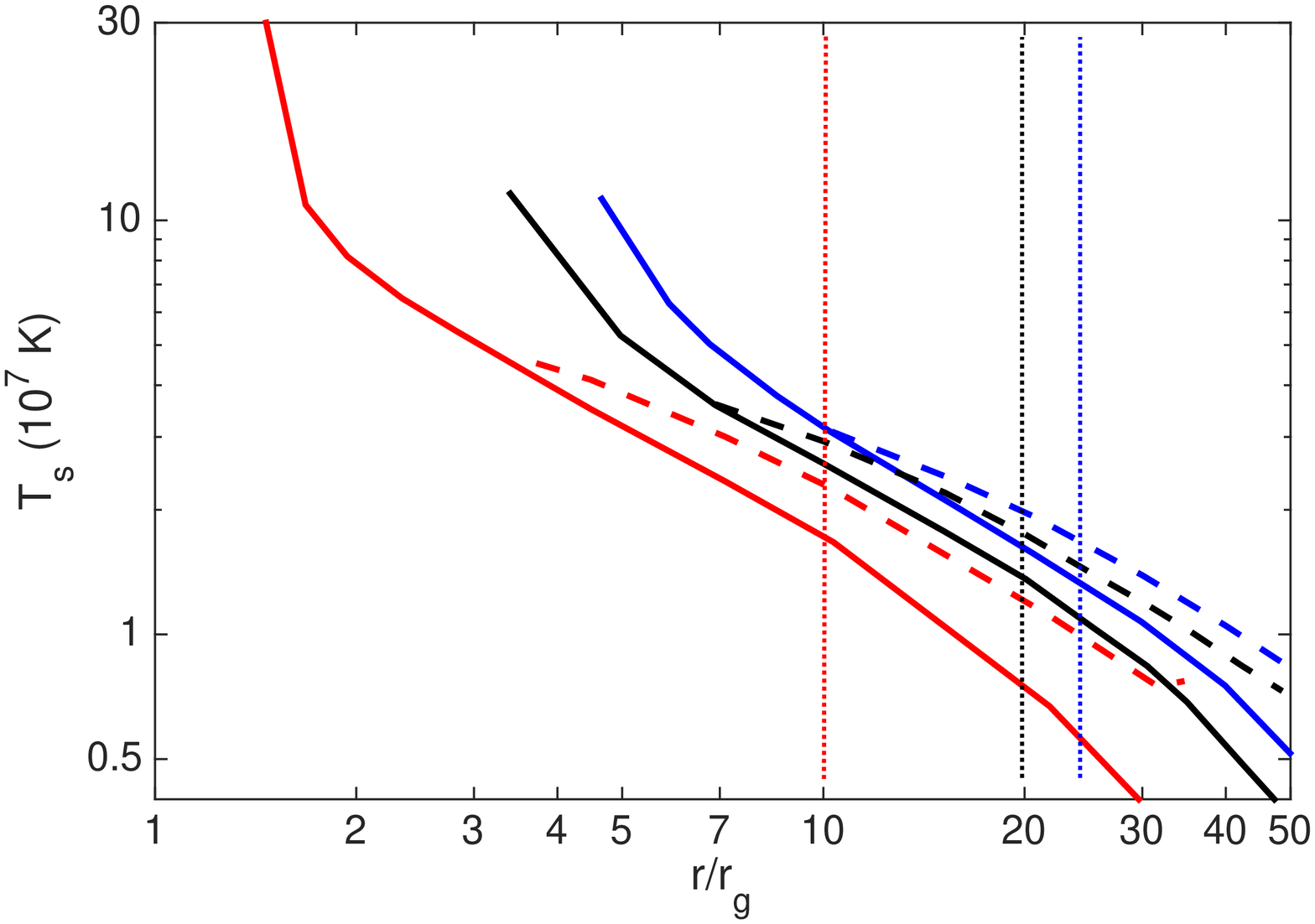}
\caption{Representative gas temperatures at the scattering (solid) and effective (dashed) photospheres for models with $a/M=0.4$ (blue), $0.7$ (black) and $0.99$ (red). The dotted vertical lines denote the radius where the total effective optical depth is approximately $10$. For the $a/M=0.4$ and $0.7$ disks, the temperature rise less steeply towards the black hole compared to that from one-zone models that neglect vertical structure \citep{ak00, db14}.}
\label{fig:tphoto}
\end{figure}

These high-spin models also exhibit stronger relativistic effects as the viewing angle changes from face-on to edge-on. In particular, when viewed edge-on the $a/M=0.99$ disk shows an energetic tail in the tens to hundreds of $\rm{keV}$ range. Despite having a photon index $\Gamma$ that falls within observed SPL range, this feature does not resemble existing data because the observed spectra generally peak at close to or less than about $10 \ \rm keV$ (see for example, GRO J1655 - 40, as shown in Figure $2$ of \cite{mr06}), whereas the blue curve in the lower panel of Figure \ref{fig:inthigh} peaks at above $30 \ \rm keV$. In this respect, the one-zone models of DB14 in fact produce a spectrum that better match observations (see Figure $1$ in their paper). On the other hand, numerical considerations placed an upper limit of approximately $200 \ \rm keV$ in our models, and it is possible that the actual disks with the same underlying physical conditions would have a broader power-law tail that extend to even higher energies, at least at certain viewing angles. It is also, however, possible that increasing the upper photon energy bound will reveal that this tail does not resemble a power-law for energies much higher than the upper limit of the plot.

More generally, none of our full-disk spectra display powerful power-law tails regardless of black hole spin. DB14 proposed that disks with non-zero torque acting at the inner boundary could naturally produce spectra resembling the steep power law (SPL) state observed in some black hole X-ray binary systems. Their argument relied on the fact that the gas temperature, which in their one-zone vertical structure approximation controls the spectral shape of individual annuli, rises sharply as distance to the black hole decreases. On the other hand, our numerical results indicate that while the disk integrated spectrum is non-thermal, the peaks of the annuli spectra are not sufficiently separated for radii near the black hole (within about $10r_{\rm g}$) to give rise to SPL-like spectra. 

We further illustrate this result by computing the gas temperature at the scattering photosphere, $T_{s}$, which approximately gives the upper limit to up-scattered photon energies. Figure \ref{fig:tphoto} shows that $T_s$ does not rise as sharply as the one-zone gas temperature in \cite{db14}. Moreover, even in the $a/M=0.99$ models where $T_s$ increases significantly more rapidly towards the black hole than in models with lower spins, the range of radii covered by the most steep rise is rather narrow and covers less than one $r_g$, as opposed to several $r_g$'s as found by DB14. This means that in our models, the hottest annuli (as measured by $T_s$) are not only not sufficiently separated in the energies where their spectra peak, but also contribute fractionally less power to the full-disk spectra, compared to those in one-zone calculations.

\section{Possible Mechanism for High-frequency Quasi-periodic Oscillations}

While our annuli spectra are not sufficiently separated in energy to give rise to SPL-like spectra, it is nonetheless still the case that the highest energy photons primarily come from the inner disk within few $r_g$'s of the ISCO. This means that the power spectra of the light curve restricted to photon energies greater than a few $\rm keV$'s would mainly reflect variability from with a narrow range of effectively thin annuli near the black hole. If such variability are due to underlying fluid oscillations determined by space-time properties, the associated frequencies would also cover only a narrow range, giving rise to sharp power spectra peaks characteristic of HFQPOs (DB14). Such a scenario would naturally explain why HFQPOs are observed in the hard X-ray bands (greater than about $2 \ \rm keV$). 

\begin{figure}
\includegraphics[width=12cm]{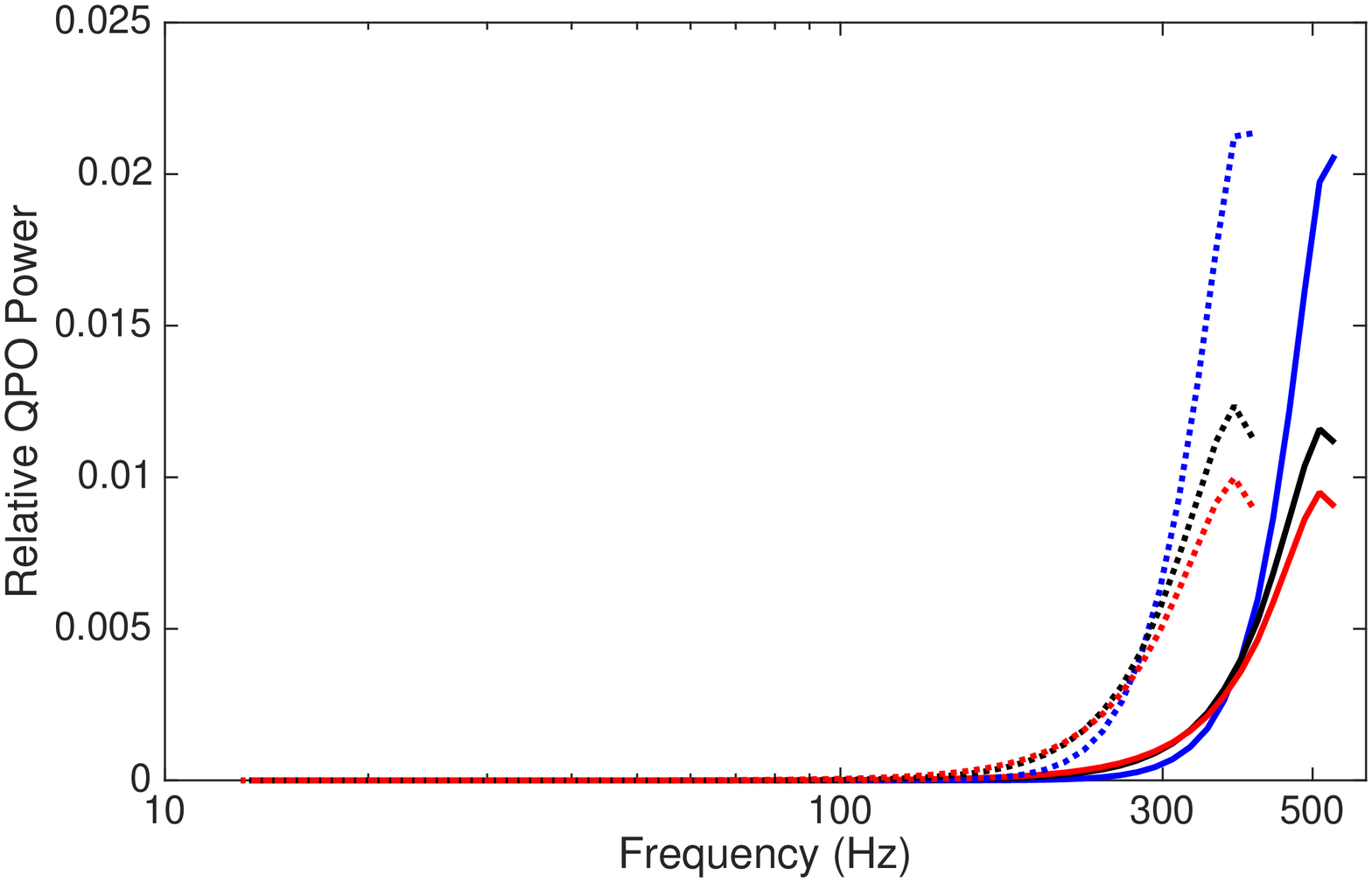}
\caption{Synthetic QPO profiles for various energy band from the $a/M=0.4$ (dotted) and $0.7$ disks (solid). The blue, black and red curves correspond to energy bands of $13 - 30 \ \rm keV$, $6 - 30 \ \rm keV$ and $2 - 30 \ \rm keV$, respectively. The horizontal axis is the local vertical epicyclic oscillation frequency at various annuli radii for the particular black hole spins.}
\label{fig:qpo}
\end{figure}

To extract potential QPO profiles from our models, we assume that each annulus oscillates at the vertical epicyclic frequency at that radius, since this mode has been observed to be prominent in both local \citep{hkb09,bla11} and global \citep{rm09} MHD simulations. To compute the relative power, we normalize the flux in each energy band at a particular radius by the total disk flux in that band, and then squaring to convert to power. Figure \ref{fig:qpo} shows representative QPO power spectra, which are prominent and have $Q$ values between approximately $2$ and $3$. These quality factors are broadly comparable to those found by DB14, except in the $13 - 30 \ \rm keV$ band where our results are smaller by about a factor of $2$. The existence of prominent QPO profiles in our models contradicts observations since the entire sample of HFQPO above $100 \ \rm Hz$ were detected in the SPL state \citep{mr06}, which do not appear in our spectra. It is plausible including additional physics that we are missing here can lead to disk models including inner torque that exhibit both SPL and HFQPOs. 

\section{Discussions and Conclusions}

We performed detailed spectral and structure calculations for accretion disks with non-zero stresses near the inner orbit. We self-consistently coupled the radiative transfer and vertical structure equations at each annulus. We also incorporated physically motivated vertical energy dissipation profiles obtained by averaging results from three-dimensional local shearing box simulations \citep{hkb09} horizontally and in time. Finally, we computed disk integrated spectra as seen by distant observers via a fully relativistic transfer function \citep{agol97}. 

Our results agree qualitatively with those based on one-zone models that essentially neglect vertical structure. However, the quantitative differences between our findings and previous work is sufficiently large to result in significantly altered disk integrated spectra. In general, the full disk spectra is non-thermal but do not display features resembling a powerful SPL tail. This is because for annuli close to the black hole ($r/r_g$ less than about $10$), the gas temperature at the scattering photosphere, which tracks the energy at which the spectrum of a saturated Compton scattering dominated atmosphere peaks, does not rise as rapidly towards the black hole as predicted by one-zone models. In other words, the spectral peaks of the effectively thin annuli are not sufficiently `separated' to form a power law tail.

Models with higher $a/M$ ($0.8$ and $0.99$) further results in even higher annuli effective temperatures near the black hole as well as stronger relativistic effects. In particular, when viewed nearly edge-on the $a/M=0.99$ disk displayed strong non-thermal tail that extended to our upper energy limit. We caution that this finding is not directly comparable the spectra generated by DB14, which exhibited powerful power-law tails regardless of viewing angle.

While our calculations suggest that including torque at the inner disk edge do not result in SPL-like spectra, we did see evidence that such models may produce HFQPOs, albeit with lower qualify factors than those from the one-zone models of DB14. This is because photons beyond about $10 \ \rm keV$ from most of our models still come primarily from a small range of annuli close to the black hole, so that the power spectra restricted to such high-energy hands contains only a narrow range of vertical epicyclic oscillation frequencies. 

We suspect that higher values of $\Delta\ep$ would produce still harder spectra with more prominent high-energy tails at lower viewing angles, which may result in SPL-like spectra with broad and energetic power-law tails that began at lower energies. However, such atmospheres are even more effectively thin, which coupled with lower surface density make obtaining converged spectral and vertical structure models challenging. These calculations are currently underway, and will appear in a follow up paper on the dependence of spectra on a wider parameter range.

As mentioned earlier, in adopting the \cite{ak00} solution for radial disk profile we have assumed a thin disk model, which is likely inappropriate for radiation pressure dominated disks radiating at near-Eddington luminosity. It would be interesting to incorporate inner torque physics into slim disk models \citep{ab88, sa11}, and use the resulting radial structure as starting point to compute the vertical structure and ultimately emergent spectra. 

Another potential extension of this work would be incorporating a wider range of simulation-based dissipation physics, especially in light of recent advances towards understanding energy transport and dissipation in accretion disks. In particular, global simulations of \cite{sc13} and spectral calculations of \cite{tb13} both suggested that strong dissipation in disk outer layers could result in energetic power law spectra, which combined with the inner torque physics we explored may offer a way to obtain both SPL spectra and HFQPOs within self-consistent disk structure and radiative transfer calculations.

Finally, we note that radiation pressure dominated disks may be thermally unstable \citep{le74}, as seen in radiation MHD shearing box simulations of \cite{jia13}. However, there is almost no observational evidence of such instabilities in BHBs. On the other hand, global simulations such as \cite{jia14} do not yet include the entire disk, and the ultimate non-linear saturation of the instability is still an open problem. 

The authors acknowledge the support from University of San Diego Faculty Research Grant, as well as fruitful discussions with O. Blaes, J. Dexter, C. Kishimoto and N. Storch. The authors would especially like thank the referee, whose report was particularly helpful for improving the accessibility of the manuscript to new researchers entering the field.

\end{document}